\documentclass[a4paper,12pt,aps,pra,superscriptaddress,reprint,floatfix,showpacs]{revtex4-1}

\usepackage[utf8]{inputenc}
\usepackage[english]{babel}
\usepackage[T1]{fontenc}
\usepackage{amsmath,amssymb}
\usepackage{graphicx}
\usepackage{braket}
\usepackage{amsfonts}
\usepackage{array}
\usepackage{lmodern}
\usepackage{esvect}
\usepackage{fancyhdr}
\usepackage{xcolor}
\usepackage{textcomp}
\usepackage[squaren,Gray]{SIunits}
\usepackage{natbib}
\usepackage{hyperref}
\usepackage[caption=false]{subfig}
\usepackage{lastpage}

\graphicspath{{Images/}}

\begin{document} 

\title{Generation of long-lived underdense channels using femtosecond filamentation in air}

\author{Guillaume Point}
\email{guillaume.point@ensta-paristech.fr}
\affiliation{Laboratoire d'optique appliquée - ENSTA ParisTech, Ecole Polytechnique, CNRS - 828 boulevard des Maréchaux, 91762 Palaiseau, France}
\author{Carles Mili\'{a}n}
\affiliation{Centre de physique théorique, Ecole Polytechnique, CNRS, F-91128 Palaiseau, France}
\author{Arnaud Couairon}
\affiliation{Centre de physique théorique, Ecole Polytechnique, CNRS, F-91128 Palaiseau, France}
\author{André Mysyrowicz}
\affiliation{Laboratoire d'optique appliquée - ENSTA ParisTech, Ecole Polytechnique, CNRS - 828 boulevard des Maréchaux, 91762 Palaiseau, France}\author{Aurélien Houard}
\affiliation{Laboratoire d'optique appliquée - ENSTA ParisTech, Ecole Polytechnique, CNRS - 828 boulevard des Maréchaux, 91762 Palaiseau, France}

\begin{abstract}
Using femtosecond laser pulses at 800 and \unit{400}{\nano\metre}, we characterize the formation of underdense channels in air generated by laser filamentation at the millijoule energy level by means of transverse interferometry. We find that using tight focusing conditions, filamentation generates a shock wave and that the resulting low-density channel lasts for more than \unit{90}{\milli\second}. Comparison of these results with hydrodynamic simulations using an Eulerian hydrodynamic code gives an good agreement and allows us to estimate the initial gas peak temperature at $\sim\unit{1000}{\kelvin}$. The influence of experimental parameters such as the focusing conditions for the ultrashort laser pulse, its polarization or the wavelength is studied and linked to previous characterizations of filamentation-generated plasma columns. 
\end{abstract}


\maketitle

\section{Introduction}

Filamentation is a propagation regime reached for short high power laser pulses propagating through transparent media. First, such pulses self-focus due to the optical Kerr effect until intensity becomes high enough to ionize the medium through multiphoton absorption and tunneling. From then on, a dynamic balance is established between, on the one hand, the Kerr effect and, on the other hand, non-linear absorption, plasma defocusing and diffraction, resulting in the beam maintaining a very high intensity in a thin channel of almost constant radius over several Rayleigh lengths \cite{Couairon2007}. In order to reach this peculiar propagation regime, however, pulses must have a peak power higher than a critical power $P_{cr} \sim \unit{10}{\giga\watt}$ in air at \unit{800}{\nano\metre}.

Once the laser pulse has left, it leaves in its trail a thin column of weakly ionized plasma, which recombines in less than \unit{10}{\nano\second} \cite{Tzortzakis2000}. During this process, energy from free electrons is transferred to rotational and translational degrees of freedom of air molecules. Two-photon Raman excitation of rotational states has also been shown to be a very efficient path of energy transfer from the laser pulse to the medium \cite{Stapelfeldt2003,Zahedpour2014}. Eventually, this energy is converted into heat of air molecules over a nanosecond timescale and it is mostly confined to the volume initially occupied by the filament \cite{Cheng2013}. Such a localized, fast energy deposition leads to the formation of an outward-propagating pressure wave and a central low-density channel presenting the same cylindrical geometry as the filament \cite{Tzortzakis2001,Wahlstrand2014}. The system gets back to its initial pressure over a microsecond timescale and the density hole decay is then governed by diffusion and can last for milliseconds \cite{Cheng2013}.

The formation of such low-density channels by filamentation has several interesting applications such as the formation of virtual optical waveguides \cite{Jhajj2014,Lahav2014}. These waveguides have the advantage to persist over the microsecond timescale, compared to filamentation-induced hydrodynamic plasma waveguides \cite{Wu2010}. It also plays a fundamental role in the field of filamentation-guided discharges. Indeed, the guiding effect and breakdown voltage reduction observed when using an ultrashort laser pulse to trigger high-voltage discharges is thought to result mostly from a standard, non-ionized air breakdown mechanism but at a lower density \cite{Vidal2000,Tzortzakis2001}. A better understanding and optimization of the formation of air underdense channels is then crucial for the development of applications of laser-guided discharges like the laser lightning rod \cite{Fontaine1999,Comtois2000,Schwarz2003,Forestier2012} or virtual plasma antennas \cite{Dwyer1984,Brelet2012}.

In this Article, we investigate the formation of underdense channels in air resulting from filamentation of a millijoule-level laser pulse. Air density profiles are recorded by means of transverse interferometry followed by Abel inversion. The full time evolution of low-density channels in the case of tight focusing ($f/35$) is recorded. In these conditions, we report the formation of channels lasting for more than \unit{90}{\milli\second}, which is almost enough to generate a permanent low-density channel with a \unit{10}{\hertz} repetition rate. Hydrodynamic simulations yield very good agreement with experiments, and enable us to estimate the initial maximum air temperature at \unit{1400}{\kelvin}. Influence of experimental parameters such as the focusing, wavelength and polarization of the laser pulse is also investigated. We find that energy transfer from the laser pulse to the medium is optimized using strong focusing, short wavelengths and linear polarization.

\section{Methods}

\subsection{Experimental methods}

\begin{figure}[!ht]
\begin{center}
\includegraphics[width = .48 \textwidth]{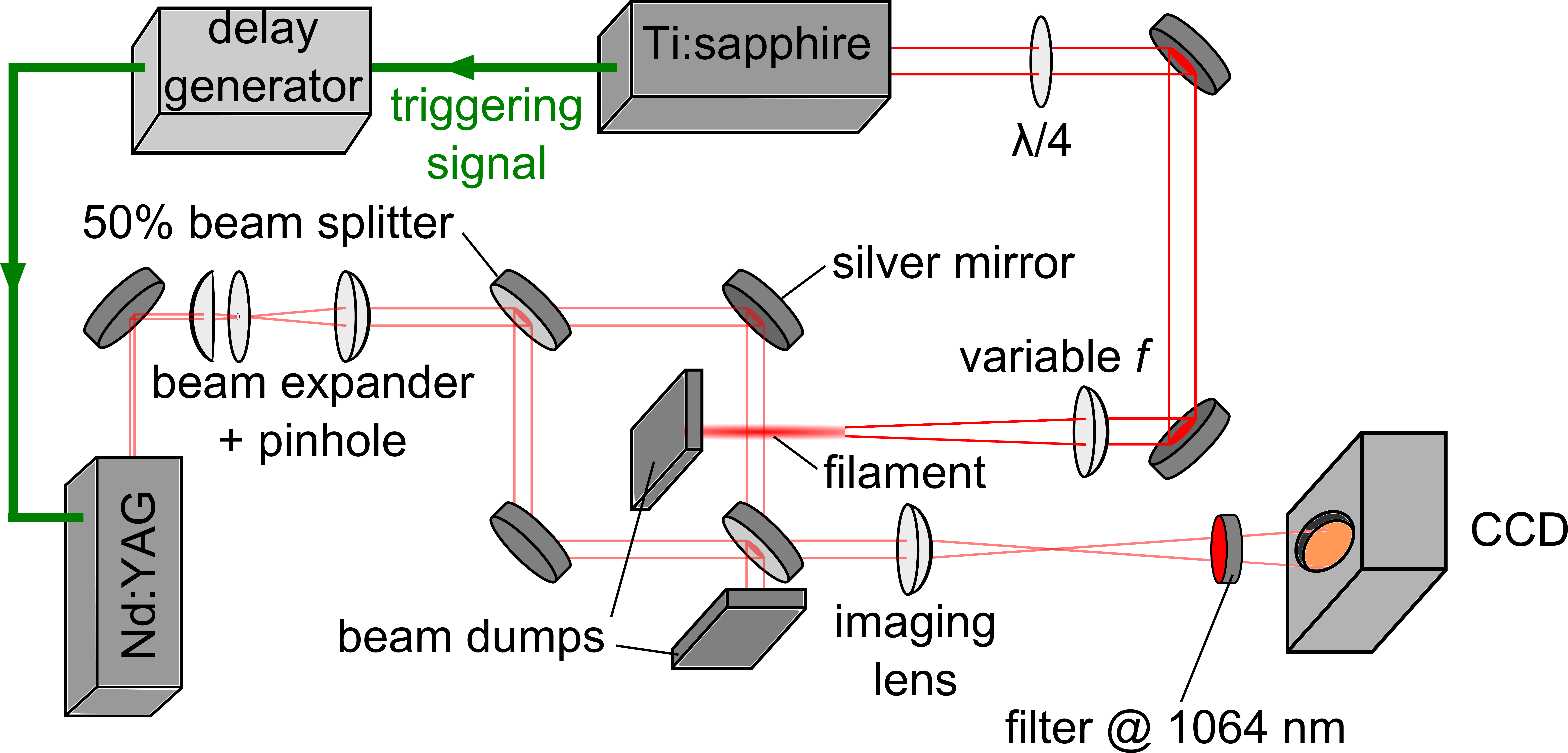}
\end{center}
\caption{Schematic representation of the experimental setup.}
\label{fig_setup}
\end{figure}

We make use of transverse interferometry to record refractive index changes induced by filamentation in air. Our interferometer is build in a standard Mach-Zehnder configuration (figure \ref{fig_setup}). The probe laser (Quanta Ray GCR-290-10 from Spectra Physics) is a Nd:YAG Q-switched laser delivering $\sim \unit{100}{\micro\joule}$, \unit{8}{\nano\second} full width at half-maximum (FWHM) pulses at \unit{1064}{\nano\metre}. The probe beam is spatially cleaned and magnified by means of a \unit{100}{\micro\metre} pinhole placed in an afocal telescope, resulting in a quasi-Gaussian spatial profile with a \unit{8.3}{\milli\metre} FWHM.

The studied underdense channels are generated by ultrashort laser pulses undergoing filamentation. The chirped pulse amplification Ti:sapphire laser chain ENSTAmobile delivers \unit{50}{\femto\second} pulses at an energy up to \unit{250}{\milli\joule} at $\lambda = \unit{800}{\nano\metre}$. For this study, we kept the pulse energy at a reference level of \unit{5}{\milli\joule}, yielding a peak power $P = \unit{100}{\giga\watt} \approx 10~P_{cr}$. A $\lambda/4$ waveplate placed at the output of the laser enables us to shift at will between linear and circular polarization. We can also make use of a second harmonic potassium dihydrogen phosphate (KDP) crystal to generate laser pulses centered at $\lambda = \unit{400}{\nano\metre}$. The pulse is focused using lenses of various focal lengths, collapsing to form a filament in one of the arms of the interferometer perpendicularly to the probe beam. A beam dump stops the filament before it reaches the reference arm, which could affect refractive index measurements. The probe laser is synchronized with the ENSTAmobile using the latter's internal clock, which grants us the possibility to set the pump/probe delay with a \unit{1}{\nano\second} precision and a \unit{1.5}{\nano\second} jitter.

\begin{figure}[!ht]
\begin{center}
\includegraphics[width = .4\textwidth]{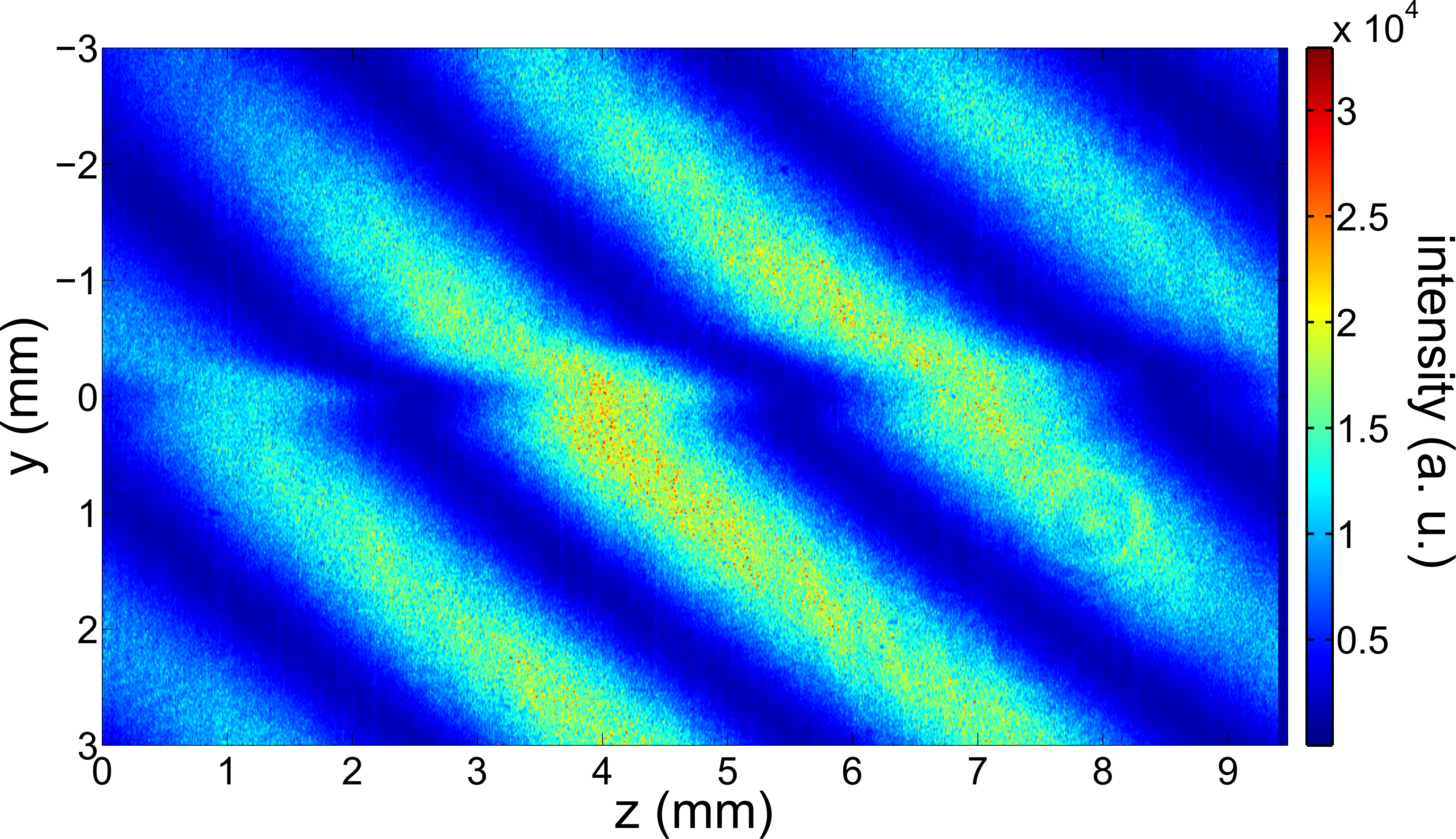}
\end{center}
\caption{Example of interferogram recorded after the filamentation of a \unit{5}{\milli\joule}, \unit{50}{\femto\second} pulse at \unit{800}{\nano\metre} focused at $f/35$.}
\label{fig_interferogram}
\end{figure}

Interferograms are recorded by the CCD camera (model TaperCamD-UCD12 from DataRay, Inc.) with a \unit{10.5}{\micro\metre} pixel size and a $1024\times1360$ pixel array, giving a field of view of more than \unit{1}{\centi\metre} in each direction. The CCD matrix is placed in the conjugated plane of the filament using a \unit{75}{\milli\metre} imaging lens and a $2f/2f$ layout.  A typical example of interferogram is given in figure \ref{fig_interferogram}. Full interferogram processing is explained in reference \cite{Point2014a}. Phase is first evaluated from the interferograms using a 1D continuous wavelet transform algorithm, and then unwrapped by a local, quality-guided phase unwrapping routine. Phase shift is obtained after removal of the background carrier phase from blank interferograms. In the single-shot regime, i. e. in the case of non-averaged data, we report a minimum RMS phase noise of \unit{2}{\milli\rad}. As the interaction length between the probe and the phase object is very short ($\sim\unit{100}{\micro\metre}$), we can completely neglect any probe deflection effect \cite{Point2014a}.

The relative dielectric constant of air, $\epsilon_{air}$, and number density $n_{n}$ are linked by the Gladstone-Dale relation, which is a simplified version of the Lorentz-Lorenz relation \cite{Born2001}:
\begin{equation}
\epsilon_{air} = \left(1+\frac{\beta}{n_{0}}n_{n}\right)^{2},
\end{equation}
where $\beta$ is the Gladstone-Dale constant. This modeling has been shown to be valid at least up to \unit{5000}{\kelvin} for air \cite{Alpher1959a}, which is far higher than typical temperatures encountered in this work. We used the empirical Ciddor equation \cite{Ciddor1996}, which links the air refractive index to pressure, temperature and humidity level, to have a precise estimation of air refractive index in the visible-near infrared range in our experimental conditions. At $p=\unit{1.013\times10^{5}}{\pascal}$ and $T=\unit{293.15}{\kelvin}$, we have $\beta = 2.7\times10^{-4}$ for a reference density $n_{0} = \unit{2.47\times10^{25}}{\rpcubic\metre}$. The recorded phase shift can then be expressed as:
\begin{equation}
\Delta\varphi = \frac{2\pi\beta}{\lambda n_{0}}\int_{s_{1}}^{s_{2}}(n_{n}(s)-n_{0})~\mathrm{d}s,
\end{equation}
where $s$ is the coordinate along the probe beam propagation axis. For a typical probed object of size $\sim \unit{100}{\micro\metre}$, we then estimate the density resolution of our interferometer to be $\sim \unit{3\times 10^{23}}{\rpcubic\metre}$, that is $\sim 0.01~n_{0}$.

Since we use a transverse geometry for our interferometric measurements, we can then recover radial density profiles using the inverse Abel transform technique. To this purpose, we use a Fourier-Hankel algorithm \cite{Smith1988}. As experimental data is never perfectly symmetric with respect to the center of the underdense channel, density is averaged between the right and the left part of inverted density profiles.

\subsection{Numerical methods}

We have solved numerically the compressible Euler equations for the fluid mass density, $\rho=n_{n} M/\mathcal{N}_A$ ($M$ being the molar mass and $\mathcal{N}_A$ the Avogadro's number), linear momentum density, $\rho\vv{v}$, and total energy per unit volume, $e$, in cylindrical coordinates neglecting azimuthal fluxes:
\begin{eqnarray}
&& \partial_tU+\vv{\nabla}\cdot F=S \\ &&
U=\left[\begin{array}{c}\rho \\ \rho\vv{v} \\ e\end{array}\right],\ 
F=\left[\begin{array}{c}\rho\vv{v} \\ \rho\vv{v}\otimes\vv{v}+p\hat I \\ \vv{v}[e+p]\end{array}\right] \\ &&
S=\left[\begin{array}{c}0 \\ \vec 0 \\ \vv{\nabla}[q\vv{\nabla} T]\end{array}\right], \ \hat I=\left[\begin{array}{cc}1 & 0\\ 0 & 1\end{array}\right].
\end{eqnarray}
Pressure, $p$, and temperature, $T$, are linked by the ideal gas equation of state (EOS) $p=\rho RT$ ($R=\unit{287.058}{\joule\cdot\kilo\rp\gram\cdot\rp\kelvin}$), and $q=\unit{2.4\times10^{-2}}{\watt\cdot\rp\metre\cdot\rp\kelvin}$ is the heat conductivity. Here $\vv{v}\otimes\vv{v}\equiv[v_r, \ v_z]^\top[v_r,\  v_z]$, where $\top$ denotes the vector transpose, and $v_{r}$ ($v_z$) is the radial (longitudinal) components of the fluid bulk velocity $\vv{v}\equiv[v_r, \ v_z]^\top$. The integration of the fluid equations is carried out by means of the Harten Lax van Leer first order Godunov method with restoration of the contact surface \cite{Toro2009}. This method is able to capture the shock wave formation, which threshold is observed to be around or below the typical peak temperatures involved in our simulations, $T\sim\unit{10^3}{\kelvin}$. The internal energy $\epsilon=e/\rho-|\vv{v}|^2/2$ and adiabatic coefficient $\gamma=1.4$ are used to compute the speed of sound, $a=\sqrt{\partial_\rho p\vert_\epsilon+p/\rho^2\partial_\epsilon p\vert_\rho}$, at each step, necessary to determine the fluxes, F. Our simulations are initialized with a Gaussian distribution of temperature. Plasma recombination occurs over a few hundreds of fs, an extremely short timescale compared to acoustic relaxation and thermal diffusion timescales ($\gtrsim$ 10 ns). We therefore assume that the heating of air occurs almost instantaneously after the laser pulse and that this yields no change in the gas density. Under this assumption the pressure and total energy are unequivocally defined and set the initial conditions.

\section{Filamentation-induced hydrodynamics}

\subsection{Experimental characterization}

\begin{figure}[!ht]
\begin{center}
\includegraphics[width = .45\textwidth]{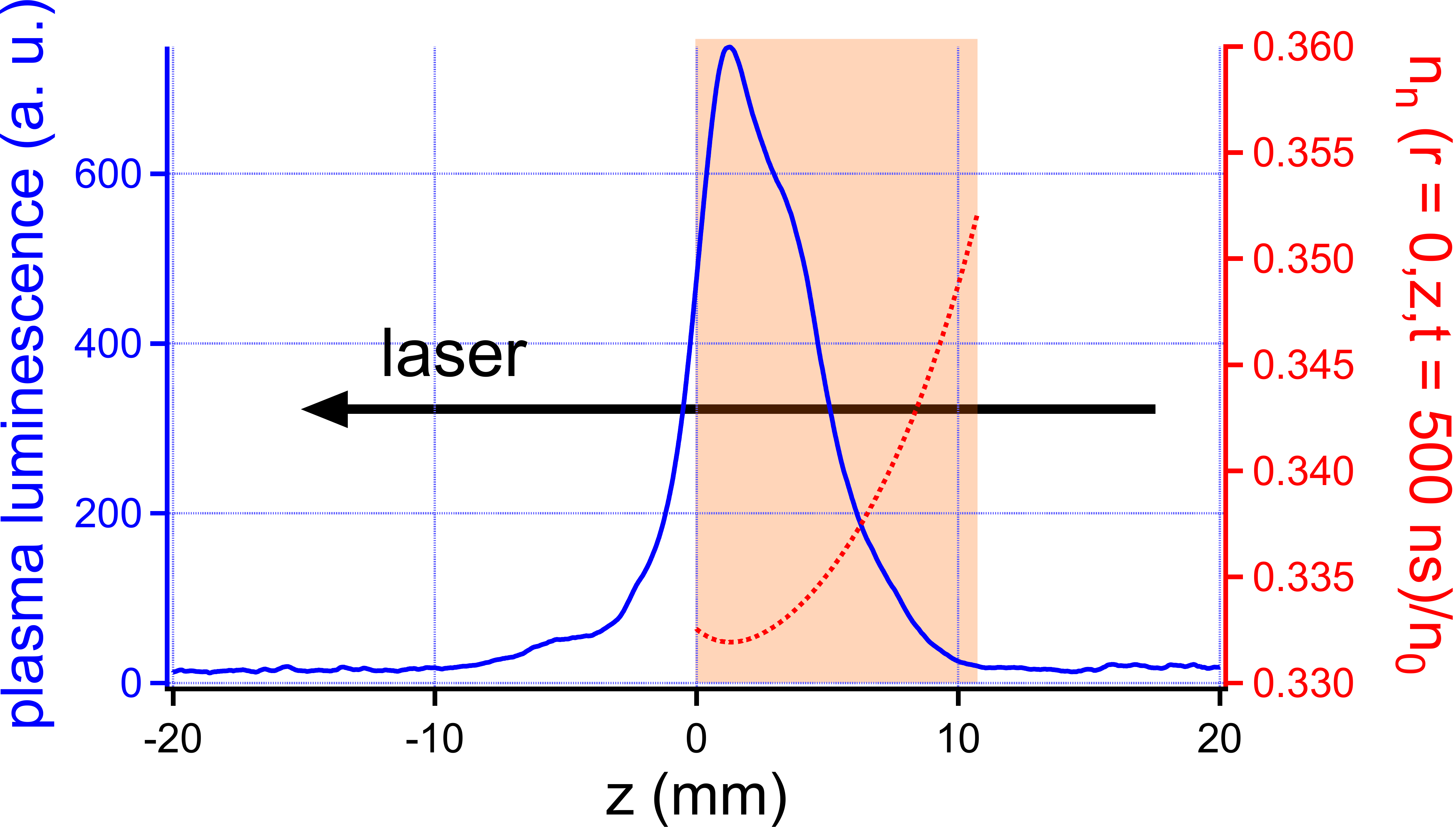}
\end{center}
\caption{$z$-evolution of the plasma luminescence (blue solid curve) and of the relative air density $n_n/n_0$ at $r = 0$ and \unit{500}{\nano\second} after the onset of filamentation of a \unit{5}{\milli\joule}, \unit{50}{\femto\second} laser pulse at \unit{800}{\nano\metre} focused at $f/355$ (red dashed curve). The area scanned by interferometry is represented by the light red rectangle.}
\label{fig_luminescence}
\end{figure}

To generally characterize the formation of underdense channels in air by filamentation, we study filaments generated by \unit{5}{\milli\joule}, \unit{50}{\femto\second}, linearly polarized pulses at \unit{800}{\nano\metre} focused at $f/35$. In this case, laser pulses collapse to form a single filament about \unit{2}{\centi\metre} long. We verified that we are indeed in the filamentation propagation regime by comparing this length, evaluated from the plasma luminescence profile (figure \ref{fig_luminescence}), with the Rayleigh length of our laser. To this purpose, we strongly attenuated the laser pulse as to be in the linear propagation regime. In the same focusing conditions, we recorded a quasi-Gaussian focal spot with waist $w_0 \sim \unit{26}{\micro\metre}$, resulting in a Rayleigh length $L_R = \unit{2.7}{\milli\metre}$, which is indeed far shorter than the plasma length.

\begin{figure}[!ht]
\begin{center}
\subfloat{\includegraphics[width=.43\textwidth]{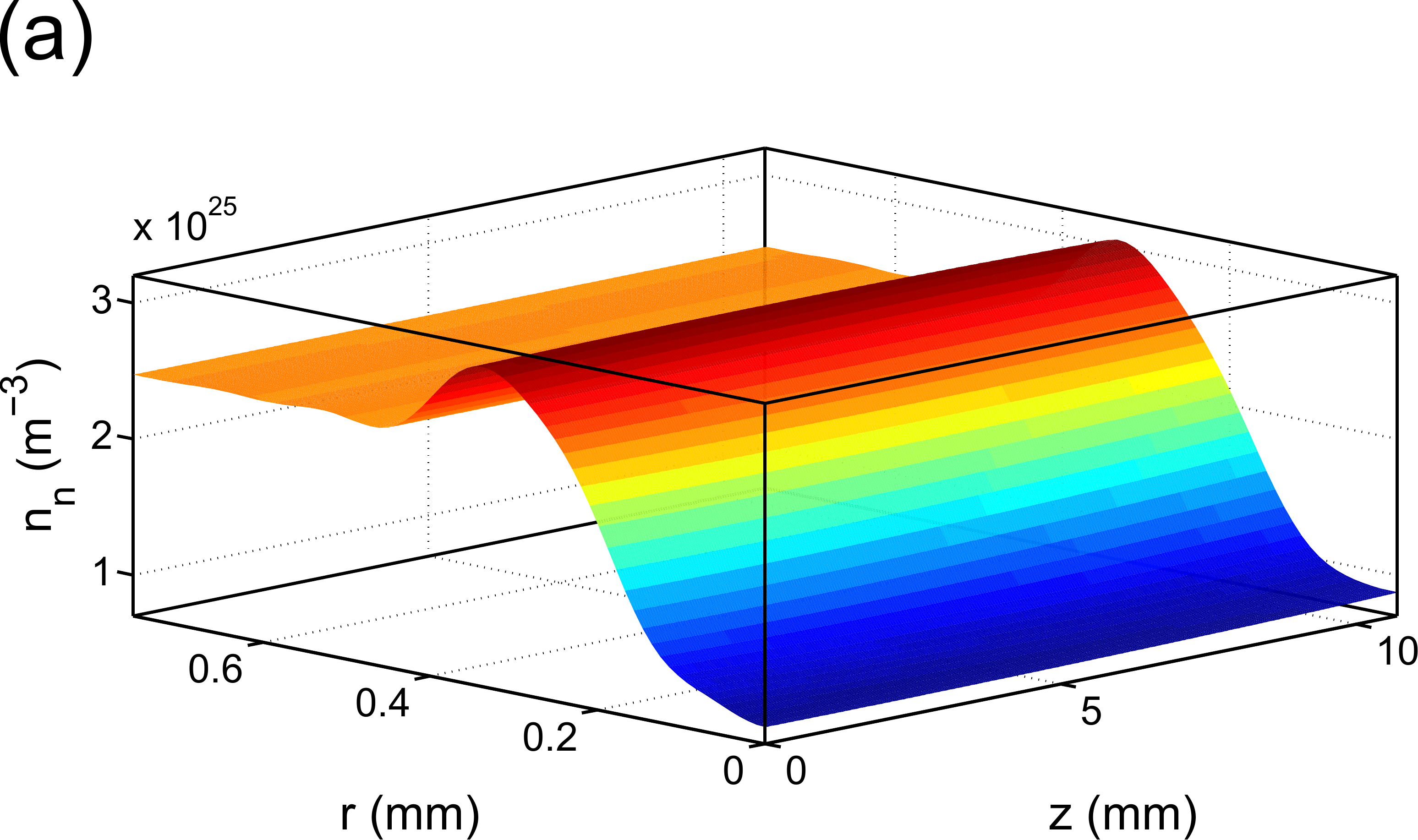}}\\
\subfloat{\includegraphics[width=.43\textwidth]{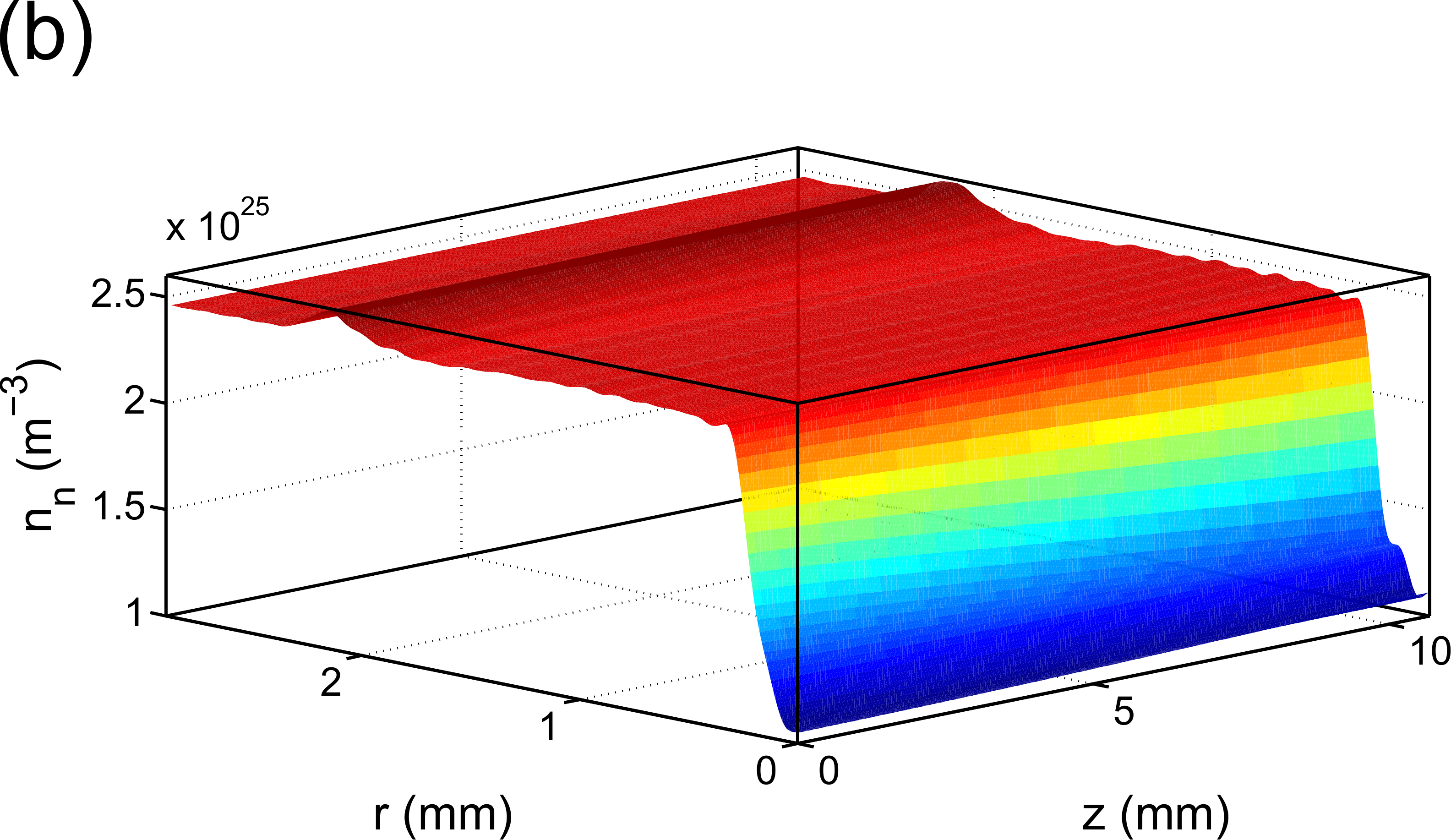}}\\
\subfloat{\includegraphics[width=.43\textwidth]{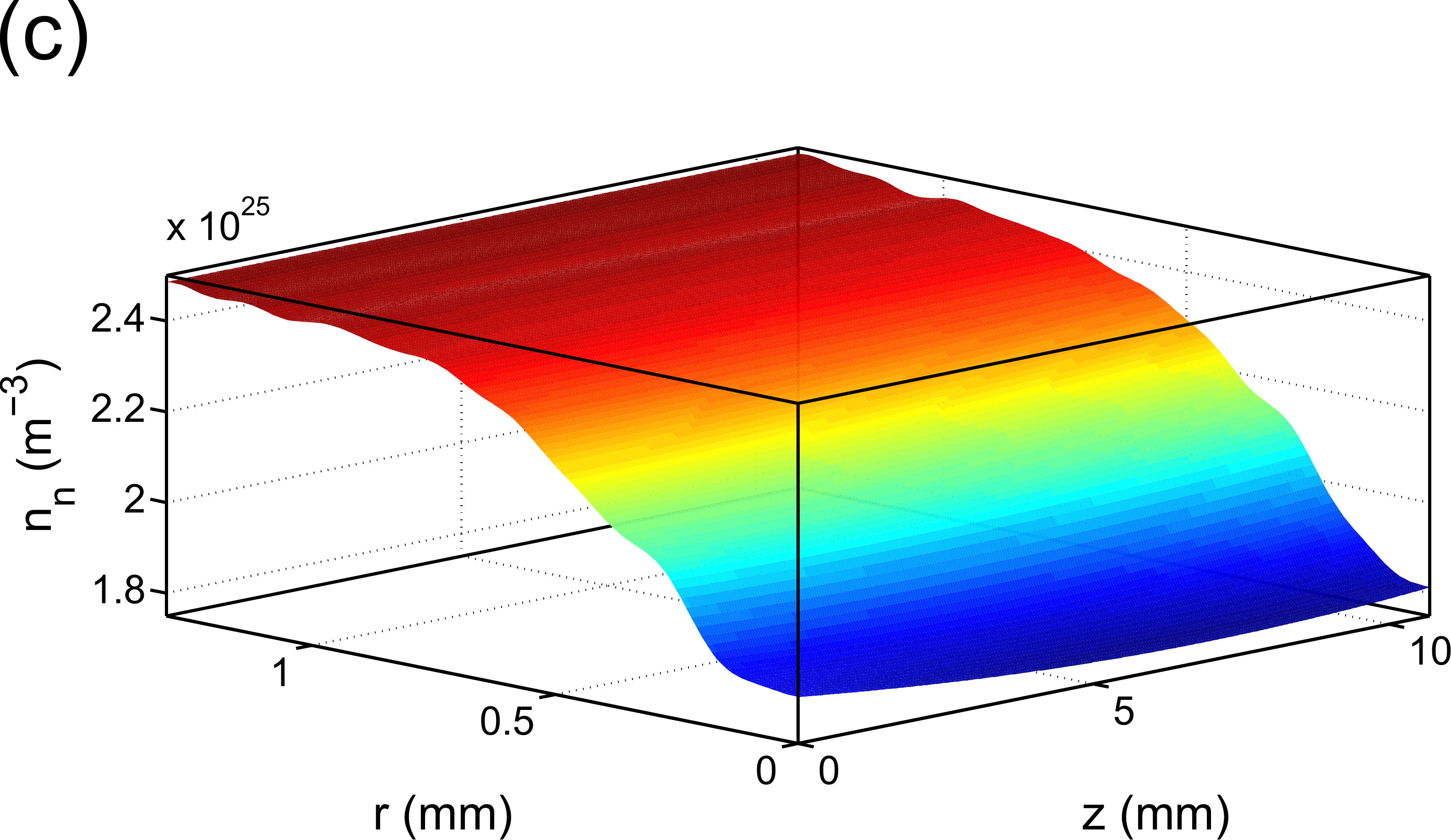}}
\end{center}
\caption{Two-dimensional $(r,z)$ neutral density profiles resulting from the filamentation of a \unit{5}{\milli\joule}, \unit{50}{\femto\second} laser pulse at \unit{800}{\nano\metre} focused at $f/35$ at delay \unit{500}{\nano\second} (a), \unit{5}{\micro\second} (b) and \unit{500}{\micro\second} (c).}
\label{fig_density_planes}
\end{figure}

The probing arm of the interferometer intercepts the filament around the lens geometric focus, where the plasma luminescence is the most intense, as highlighted by the light red rectangle in figure \ref{fig_luminescence}. Interferometric acquisitions are done in the single shot regime, and no subsequent multi-shot data averaging is done. The strength of our diagnostic lies in its ability to give unambiguous two-dimensional space-resolved density profiles thanks to the Abel inversion process. In figure \ref{fig_density_planes} are presented such two-dimensional profiles for three different delays: 0.5 (figure \ref{fig_density_planes}-(a)), 5 (figure \ref{fig_density_planes}-(b)) and \unit{500}{\micro\second} after the onset of filamentation (figure \ref{fig_density_planes}-(c)). Here, $r$ corresponds to the radial coordinate and $z$ to the laser propagation axis, that is the cylindrical symmetry axis.

Figure \ref{fig_density_planes}-(a) clearly shows the formation of a central underdense channel of $\sim \unit{350}{\micro\metre}$ FWHM and of minimum value $\sim \unit{8.5\times10^{24}}{\rpcubic\metre}$, that is a $60\%$ reduction with respect to $n_{0}$. This channel is surrounded by a annular overdense ridge centered around $r=\unit{330}{\micro\metre}$ of which amplitude is $\sim \unit{3\times10^{25}}{\rpcubic\metre}$. A simple evaluation of the $n_{l}$ parameter, that is the relative linear neutral density, defined as:
\begin{equation}
n_{l}(z) = \left|\frac{\displaystyle\int_{\mathbb{R}^{+}}(n_{n}(r,z)-n_{0})r~\mathrm{d}r}{\displaystyle\int_{\mathbb{R}^{+}}n_{0}r~\mathrm{d}r}\right|
\end{equation}
gives $\displaystyle\max_{z}(n_{l}) \sim 10^{-3}$, proving the conservation of the number of neutral molecules is respected. 

Looking at the density profiles for the \unit{5}{\micro\second} delay (figure \ref{fig_density_planes}-(b)), we see the low-density channel is now shallower and wider, reaching the level of $\unit{10^{25}}{\rpcubic\metre}$ for a \unit{400}{\micro\metre} FWHM. As for the overdense ring, it has propagated outwards, being now around $r = \unit{2.2}{\milli\metre}$. We can thus roughly estimate the speed of this hydrodynamic wave at $u \sim \unit{415}{\metre\cdot\rp\second}$, that is slightly more than the speed of sound in standard pressure and temperature conditions $a = \unit{343}{\metre\cdot\rp\second}$. This wave completely attenuates and results in a negligible contribution to the phase shift after a few tens of microseconds. After \unit{500}{\micro\second} (figure \ref{fig_density_planes}-(c)), the underdense channel continues to resorb by diffusion, enlarging and decreasing in amplitude at \unit{800}{\micro\metre} and $\sim\unit{1.8\times10^{25}}{\rpcubic\metre}$ respectively.

We can see that, at all times, air density at the center of the underdense channel is not uniform along $z$. This phenomenon undoubtedly results from a inhomogeneous laser energy deposition in the medium. However, as seen in figure \ref{fig_luminescence}, the amplitude of the underdense channel at early times (\unit{500}{\nano\second}) and of the plasma luminescence are only weakly correlated, the channel extension in $z$ being obviously far larger than the luminescent channel. As luminescence is related to the presence of plasma, we can thus assert that ionization as a channel for energy transfer to the medium is very localized in space, and that rotational absorption obviously occurs long before and far after plasma generation along the beam propagation axis. This observation agrees well with the simulations recently done by Rosenthal \textit{et al.} with similar pulse peak power \cite{Rosenthal2014}.

\begin{figure}[!ht]
\begin{center}
\includegraphics[width=.45\textwidth]{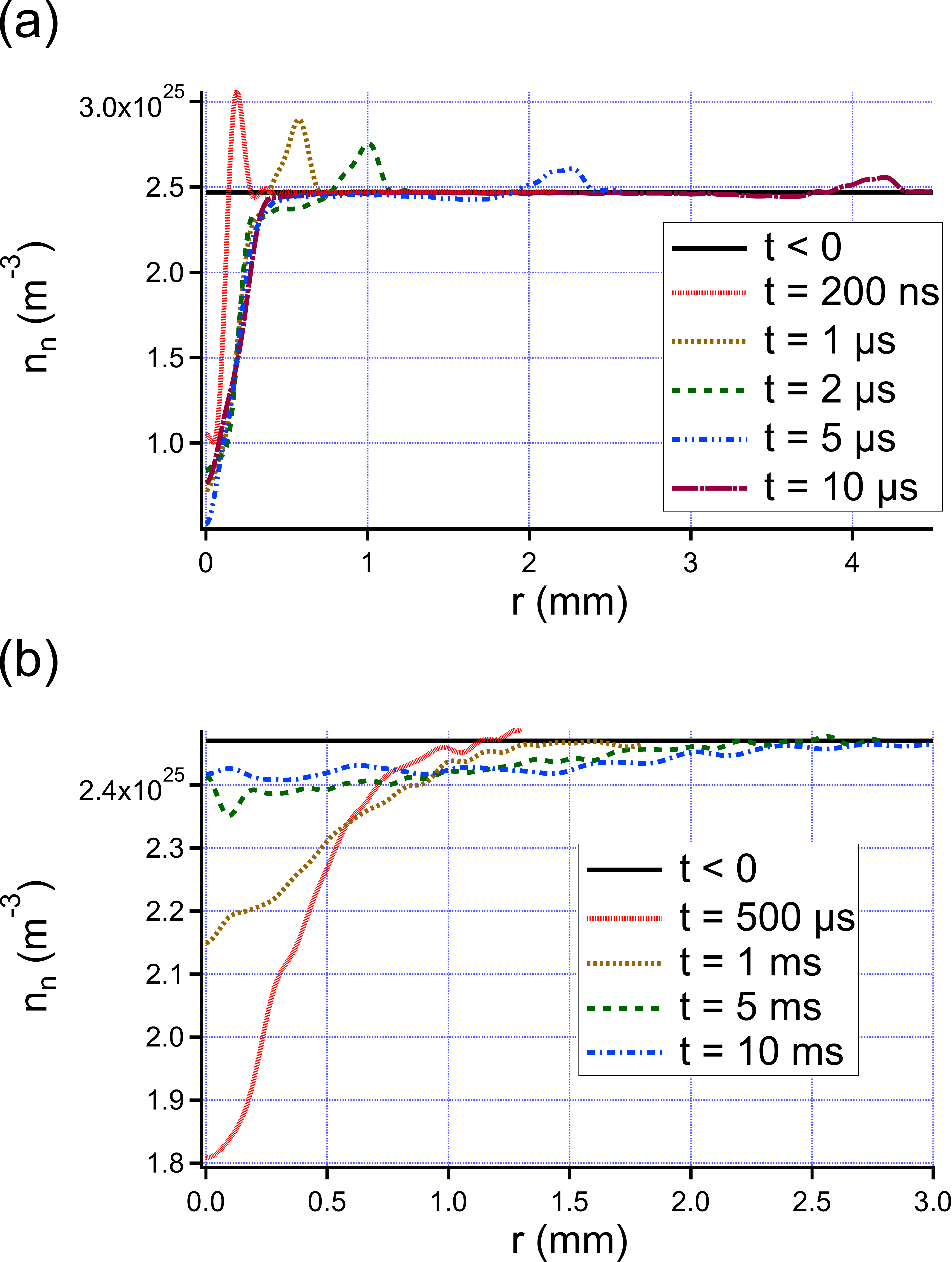}
\end{center}
\caption{$z$-averaged density profiles for short (a) and long delays (b) resulting from the filamentation of a \unit{5}{\milli\joule}, \unit{50}{\femto\second} laser pulse at \unit{800}{\nano\metre} focused at $f/35$.}
\label{fig_density_profiles}
\end{figure}

A clearer picture of the whole phenomenon can be obtained with one-dimensional density profiles. To this purpose, data from density planes is averaged along $z$, yielding the curves displayed in figure \ref{fig_density_profiles}. It is notably easier to follow the propagation of the single-cycle sound wave and its progressive attenuation with time in figure \ref{fig_density_profiles}-(a). The strong density gradients encountered in this wave and its supersonic speed are characteristic of a shock wave, a very interesting feature which was not seen in the case of less energetic laser pulses \cite{Cheng2013,Wahlstrand2014}.

Long-time dynamics (figure \ref{fig_density_profiles}-(b)) are characterized by the persistence of the underdense channel, which slowly resorbs by diffusion. We could still record a phase shift on the order of a few tens of milliradians at a delay of \unit{90}{\milli\second}, but the channel is then so large that it affects the whole field of view of the camera and prevents the Abel inversion of the profile to recover density. It might then be possible, in these conditions, to witness the appearance of a cumulative effect and the formation of a permanent low-density channel with a laser frequency of only \unit{10}{\hertz}, much in the same way it was observed by Cheng \textit{et al.} for low energy laser pulses at a \unit{1}{\kilo\hertz} cadence \cite{Cheng2013}.

We have been able to fully characterize filamentation-induced hydrodynamics in air by means of a two-dimensional space resolved diagnostic. We now can investigate the influence of different experimental parameters on these effects. For the sake of clarity, we will only be using $z$-averaged density profiles, resulting in 1D data sets. 

\subsection{Numerical reconstruction}
\label{sec_num_recons}

Two-dimensional simulations of the experimentally observed cylindrical geometry showed identical results as the 1D case (neglecting axial dynamics). We therefore show the latter case for the sake of transparency. Simulations were initialized using a Gaussian temperature profile, yielding two variable parameters: maximum initial temperature $T_{max}$ and spatial standard deviation $\sigma$. The best fit overall between numerically computed profiles and experimental profiles over the first \unit{10}{\micro\second} was obtained for an initial temperature profile of amplitude \unit{1400}{\kelvin} and of standard deviation $\sigma = \unit{85}{\micro\metre}$. By varying the width in about $\pm \unit{15}{\micro\metre}$ and the peak temperature by $\pm \unit{200}{\kelvin}$, the overall agreement worsens significantly in terms of the sound wave position and amplitude.

\begin{figure}[!ht]
\begin{center}
\includegraphics[width = .45\textwidth]{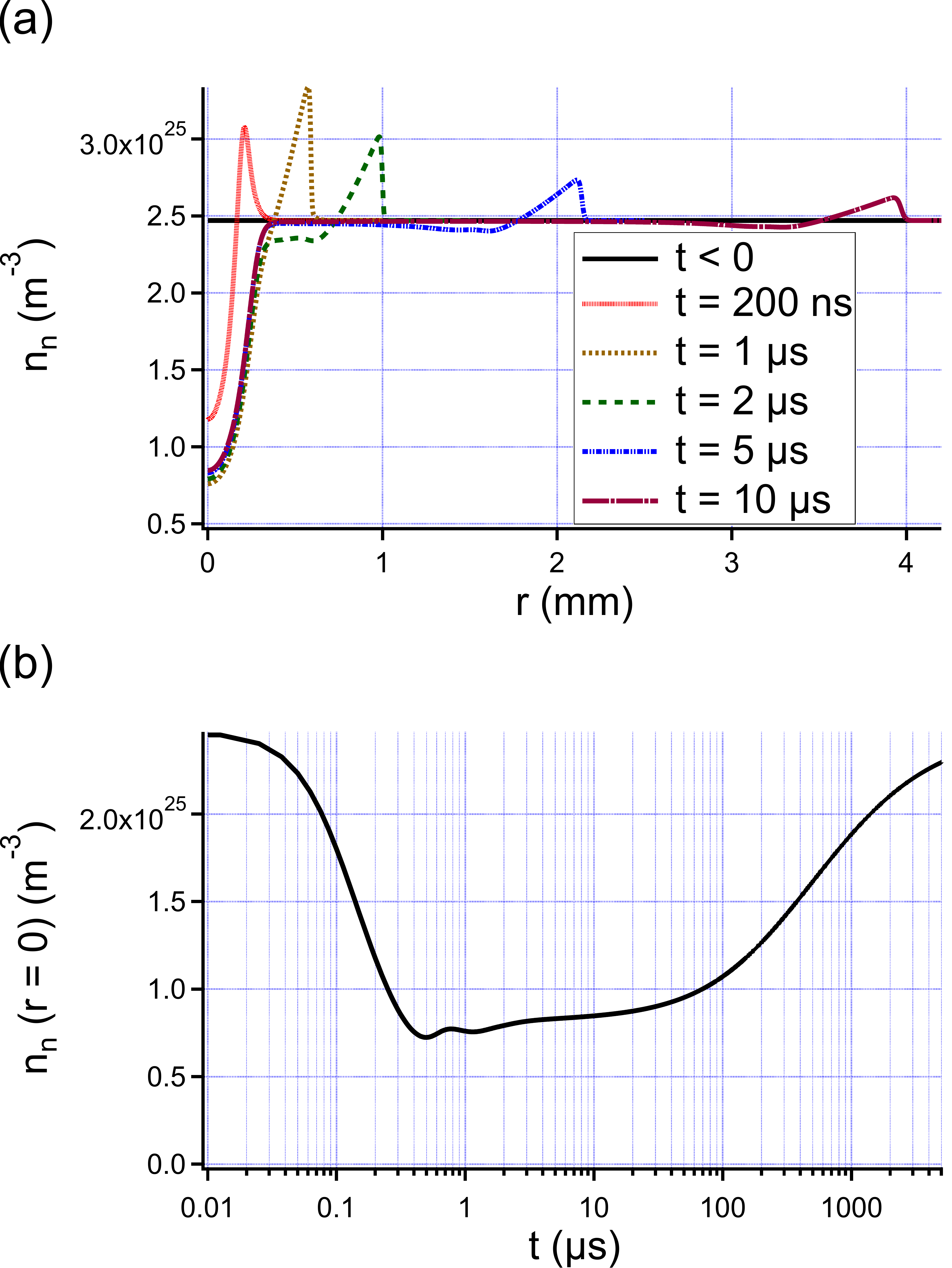}
\end{center}
\caption{
(a): density profiles for the same times as in figure \ref{fig_density_profiles}-(a) from the best fit simulation. (b): evolution of the on-axis density with time from the best fit simulation shown in (a).}
\label{fig_simu}
\end{figure}

Figure \ref{fig_simu}-(a) displays five different density profiles from the best fit taken at the same times as in figure \ref{fig_density_profiles}-(a). The strong central heating results in the formation of a supersonic shock wave and of a central underdense channel in a very similar way as witnessed in the experiment. The position of the wave is in excellent agreement in the two cases. Moreover, the peak and minimum neutral densities here encountered are also very close to their experimental counterparts, at about $\unit{3\times10^{25}}{\rpcubic\metre}$ and $\unit{8\times10^{24}}{\rpcubic\metre}$ respectively. Also, as seen in figure \ref{fig_simu}-(b), which displays the time evolution of on-axis density, the minimum density is reached around \unit{1}{\micro\second}, corresponding to what was actually seen experimentally. After this point, density goes up very slowly, eventually reaching $\unit{2.3\times10^{25}}{\rpcubic\metre}$ after \unit{5}{\milli\second}.

To verify the physical validity of an initial temperature on the order of \unit{1000}{\kelvin}, we measured the total deposited laser energy in the medium, yielding $\unit{780 \pm 160}{\micro\joule}$ (measured over 500 shots). Considering that heat is released in a isochore fashion, we can then link the variation of energy of the system to the variation of temperature through the first law of thermodynamics:
\begin{equation}
\Delta E = c_v N \Delta T,
\end{equation}
where $c_v$ is the isochore molar heat capacity of air and $N$ the number of involved air molecules. If we consider the limit case for which all the deposed energy is converted into heat (that is, we neglect radiative losses and energy storage in the internal degrees of freedom of air molecules), using a hot channel of length $\sim \unit{2}{\centi\metre}$ and of radius $\sim \unit{120}{\micro\metre}$ (corresponding to what was used in the simulation), we find:
\begin{equation}
\Delta T \sim \unit{1000}{\kelvin}.
\end{equation}

In summary, the simulation is in excellent agreement with experimental results and thus we can estimate the initial temperature induced by the laser energy deposition in air to be about \unit{1400}{\kelvin}. Note this is achieved with multi-millijoule pulses and the heating is one order of magnitude higher than previously estimated temperatures achieved with $\sim\unit{100}{\micro\joule}$ \cite{Cheng2013}.

\section{Influence of experimental configuration on underdense channels}

\subsection{Influence of focusing conditions}

External focusing conditions of an ultrashort laser pulse have been shown to have a strong impact on the resulting filament. Moderate to strong focusing (up to $\sim f/60$) results in a gentle enlargement of the filament core and a rise of the plasma density \cite{Theberge2006}. Under stronger focusing conditions ($\sim f/10$), it is even possible to reach very high plasma densities up to the optical breakdown of air \cite{Kiran2010a,Liu2010}. This is attributed to the increase of laser energy at the center of the beam due to the lens-imposed phase which, in turn, increases peak intensity and plasma density. As a new dynamic balance between focusing, self-focusing, non-linear absorption and plasma defocusing is reached, the intense core of the filament swells to compensate for the increase in electron density.

\begin{figure}[!ht]
\begin{center}
\includegraphics[width=.4\textwidth]{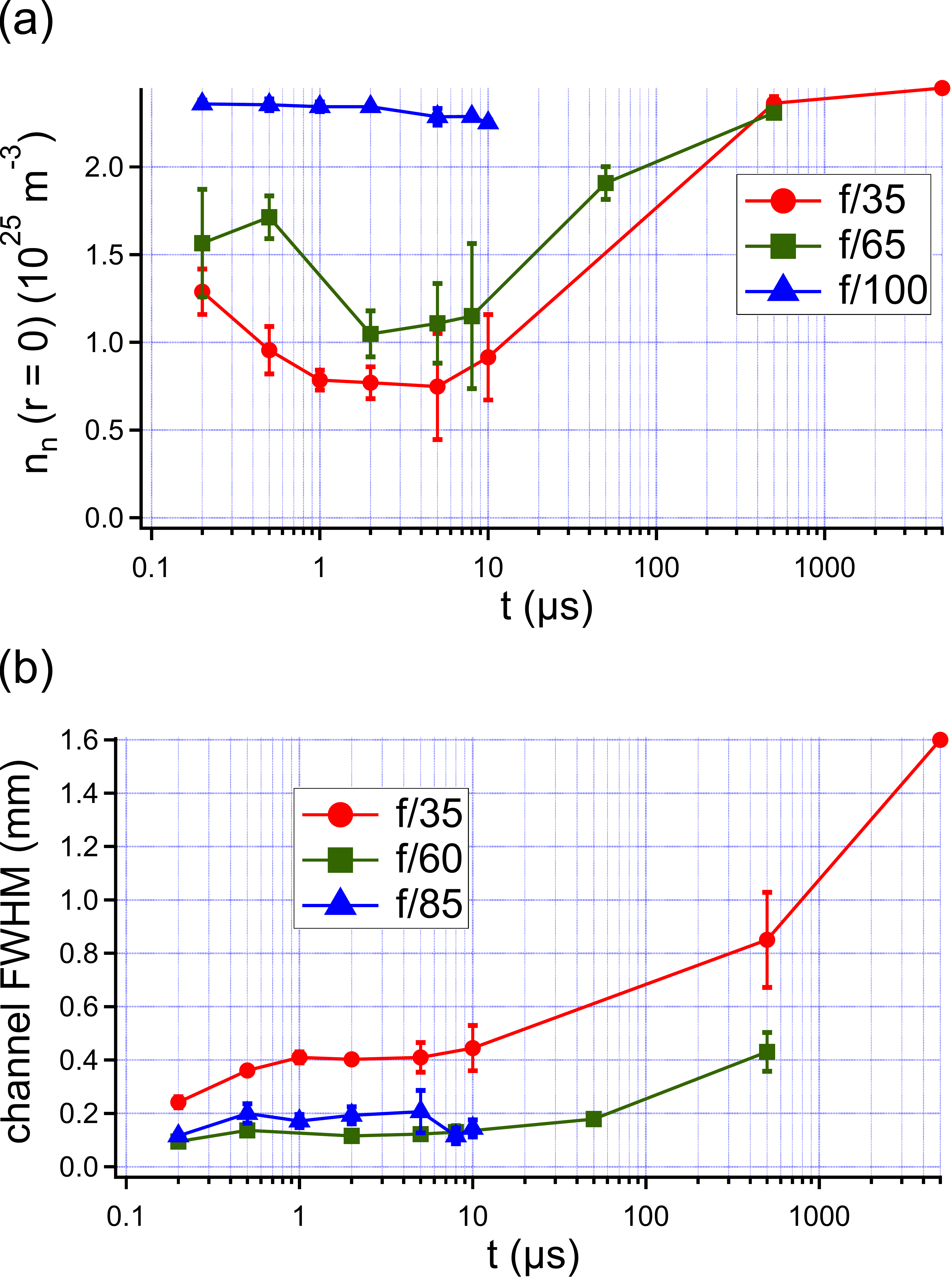}
\end{center}
\caption{Evolution of the on-axis neutral density (a) and of the channel FWHM (b) with time for different focusing conditions for a \unit{5}{\milli\joule}, \unit{50}{\femto\second} pulse at \unit{800}{\nano\metre}.}
\label{fig_lens}
\end{figure}

To study the influence of the external focusing on filamentation-generated underdense channels, we followed the evolution of density profiles with time in the case of moderate ($f/85$), strong ($f/60$) and very tight focusing ($f/35$). Corresponding results are plotted in figure \ref{fig_lens}. Figure \ref{fig_lens}-(a) shows the time evolution of the on-axis air density, that is the depth of the underdense channel, while figure \ref{fig_lens}-(b) gives an estimation of the radial extension of the channel in time. Laser energy deposition was also measured for each case, giving $\unit{570 \pm 150}{\micro\joule}$ for the $f/85$ case, $\unit{590 \pm 130}{\micro\joule}$ for the $f/60$ case and $\unit{780 \pm 160}{\micro\joule}$ for the $f/35$ case (statistics over 500 shots). It appears that the stronger the focusing, the deeper and wider the low-density channel. More specifically, we notice that at first, the channel width remains almost constant while its depth increases dramatically. As the focusing is further tightened, the underdense channel marginally deepens but, this time, widens significantly.

These results are a direct consequence of the evolution of plasma density with different focusing conditions. Indeed, heating of air arises mainly from two different effects: the first one is the Raman rotational excitation of air molecules, and the other one the conversion of plasma electron kinetic energy to molecular rotational, vibrational and translational thermal energy. Since peak intensity varies slowly between tight and moderate focusing conditions due to the effect of plasma defocusing and multiphoton absorption, the only vector for energy transfer from the laser pulse to the medium that is strongly affected by the focusing parameter is plasma generation. 

At first, deposited laser energy does not vary significantly between the $f/85$ and $f/60$ cases. However, plasma luminescence was measured to be only \unit{6}{\centi\metre} long in the latter case, to be compared with \unit{12}{\centi\metre} in the former case. As a consequence, as the channel width is the same in both cases, we should witness an underdense channel about two times deeper at $f/60$ than at $f/85$. This prediction is indeed verified experimentally. Note that this phenomenon has already been witnessed by Théberge \textit{et al.} in the case of a $2.5~P_{cr}$ laser pulse, with a peak plasma density increasing two-fold between $f/120$ and $f/60$ \cite{Theberge2006}. For an even stronger focusing at $f/35$, deposited energy becomes significantly larger while the plasma length is reduced to \unit{2}{\centi\metre}. This absorption undoubtedly originates from the tremendously increased plasma generation which, in turn, raises the gas temperature to the \unit{1000}{\kelvin} level. This results in the formation of a shock wave instead of a sound wave (as seen in section \ref{sec_num_recons}) and, therefore, in an underdense channel much larger than in previous cases, but only marginally deeper.

\subsection{Influence of the wavelength}

\begin{figure}[!ht]
\begin{center}
\includegraphics[width=.4\textwidth]{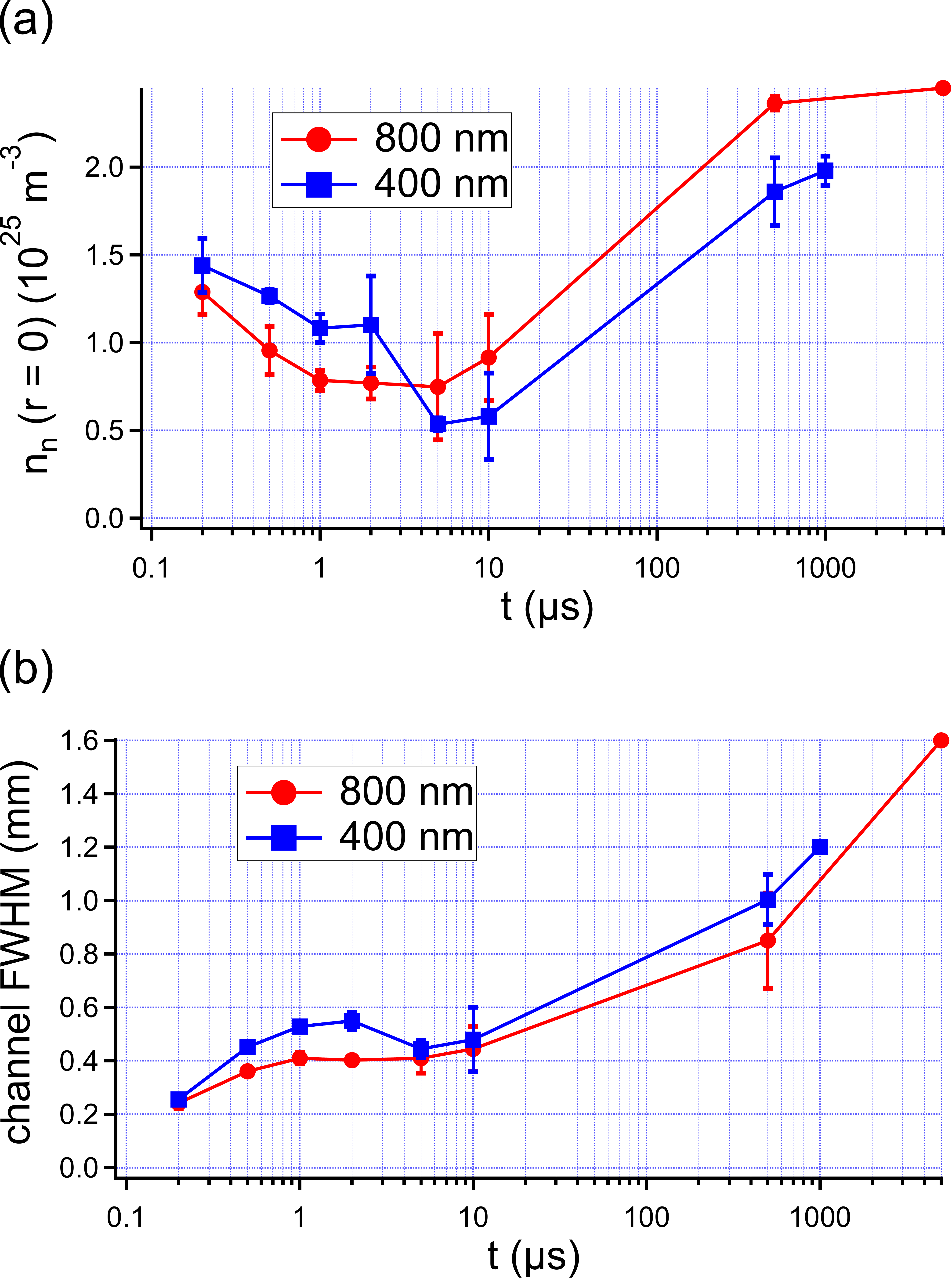}
\end{center}
\caption{Evolution of the on-axis neutral density (a) and of the channel FWHM (b) with time for a \unit{5}{\milli\joule}, \unit{50}{\femto\second} laser pulse at $\lambda = \unit{800}{\nano\metre}$ (red circles) and $\lambda = \unit{400}{\nano\metre}$ (blue squares) focused at $f/35$.}
\label{fig_wavelength}
\end{figure}

The same comparative study is performed for \unit{5}{\milli\joule}, \unit{500}{\femto\second} pulses focused at $f/35$ at both 800 and \unit{400}{\nano\metre}. The corresponding results are displayed in figure \ref{fig_wavelength}.

As it can be seen in \ref{fig_wavelength}-(a), blue filaments are responsible for the generation of a slightly deeper underdense channel than in the case of infrared filamentation, which however takes a longer time to develop. This channel also appears larger in the \unit{400}{\nano\metre} case (figure \ref{fig_wavelength}-(b)). We report a minimum density level of $\unit{4\times10^{24}}{\rpcubic\metre}$ reached at \unit{10}{\micro\second}, that is a density reduction of more than $75\%$ with respect to $n_{0}$.

These results are in good agreement with what was observed by Zhang \textit{et al.} \cite{Zhang2009b}. They characterized the filamentation of focused and collimated \unit{6}{\milli\joule}, \unit{170}{\femto\second} pulses at \unit{400}{\nano\metre}, which almost corresponds to the same regime as ours. They found blue filaments focused at $f/35$ resulted in a $\sim\unit{160}{\micro\metre}$ wide plasma channel with a typical density $\unit{2\times10^{23}}{\rpcubic\metre}$, where infrared filaments had a FWHM of only \unit{110}{\micro\metre} in the same conditions. We can also estimate from the work of Théberge \textit{et al.} that infrared filaments obtained with similar focusing have a density around $\unit{10^{23}}{\rpcubic\metre}$ \cite{Theberge2006}. Therefore, plasma columns generated by blue filamentation are slightly larger and denser than their infrared counterpart, which corresponds to what we observed studying the resulting underdense channels.

\subsection{Influence of the polarization}

Polarization dynamics in the field of laser filamentation are not well elucidated. It is agreed that critical power in case of circularly polarized pulses is higher than for linear pulses, and that multiple filamentation is more difficult to achieve in these conditions \cite{Kolesik2001a,Fibich2003}. Numerical studies in the case of $\sim 3P_{cr}$ pulses have shown that the generated plasma is less dense for circular polarisation than for linear polarization \cite{Kolesik2001a}. However, experiments, both at filamentation threshold at higher power ($\sim 20~P_{cr})$ gave either a similar electron density in both cases \cite{Diels2010}, or a higher plasma density with circularly polarized pulses \cite{Petit2000}.

\begin{figure}[!ht]
\begin{center}
\includegraphics[width=.4\textwidth]{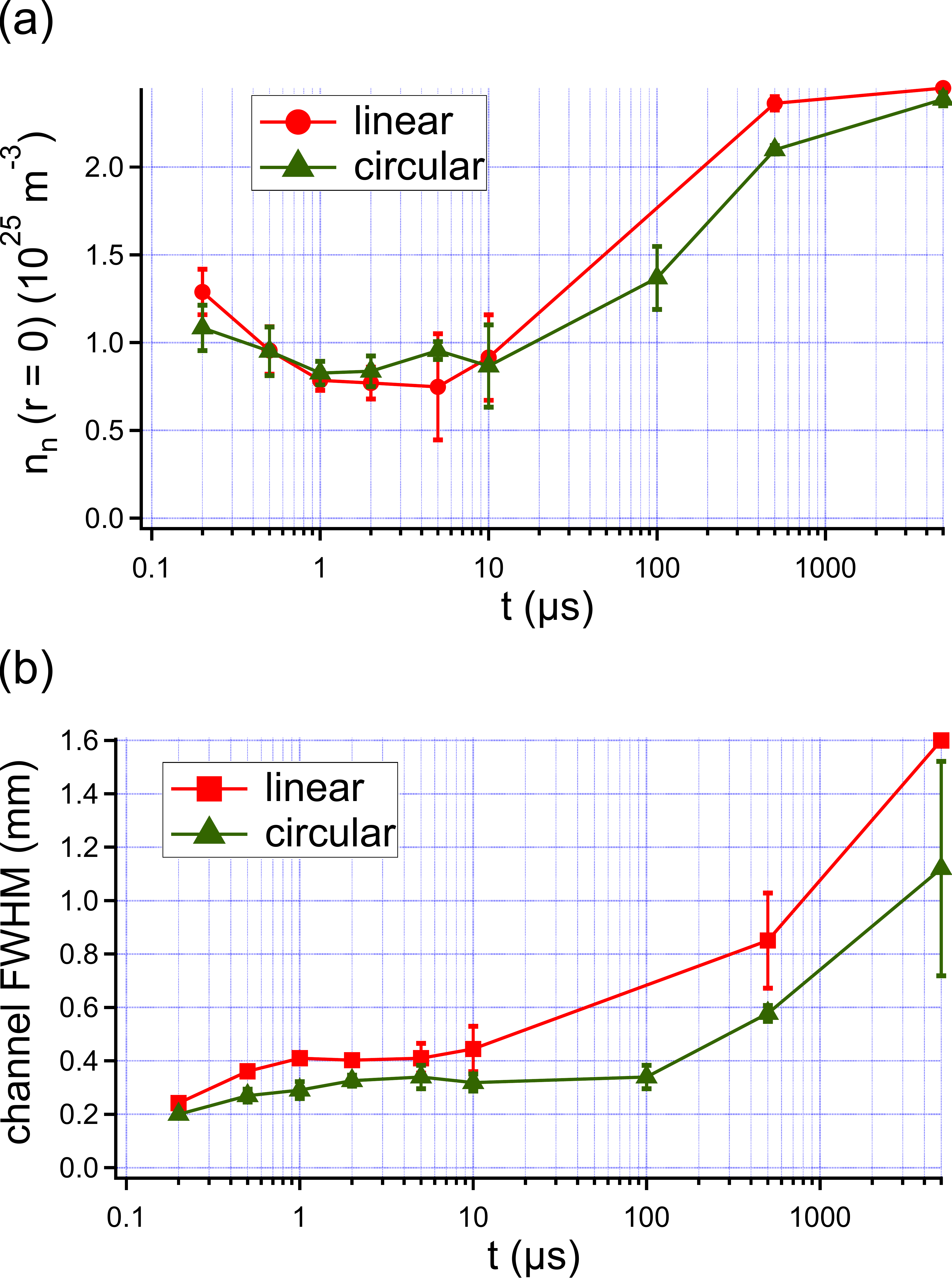}
\end{center}
\caption{Evolution of the on-axis neutral density (a) and of the channel FWHM (b) with time for a \unit{5}{\milli\joule}, \unit{50}{\femto\second} laser pulse at \unit{800}{\nano\metre} with linear polarization (red circles) and and circular polarization (green triangles).}
\label{fig_polarization}
\end{figure}

However one also has to bear in mind that energy transfer from the laser pulse to the plasma in these two different configurations is not the same. First, as evidenced by previous experimental results, the total number of electrons produced at the center of the filamentation channel for circular polarization is equal or superior to that in the case of linear polarization. In a similar volume, more laser energy is then transferred to the medium in the form of free electron potential energy for circular polarization. Moreover, it has been calculated that on average, electron kinetic energy in the case of circular polarization is well defined around a value almost proportional to the laser intensity, since electron mostly acquire energy through ponderomotive interaction \cite{Zhou2011}. This value can reach up to \unit{15}{\electronvolt}, as evidenced by the existence of collision-assisted population inversion of the triplet manifold of molecular nitrogen in plasma filaments \cite{Mitryukovskiy2014}. On the other hand, in the case of linear polarization, the electron energy distribution is peaked at low energy and weakly extends up to only a few electronvolts. As a consequence, plasma generation should then be a better vector for energy transfer from the laser pulse to the medium in the case of circular polarization.

The results of our experiments are presented in figure \ref{fig_polarization}. Figure \ref{fig_polarization}-(a), which displays the minimum neutral density in the underdense channel with time, does not show a strong difference between the two polarizations. Regarding the channel size (figure \ref{fig_polarization}-(b)), experimental observations are however different with a noticeably larger channel for the linear case than for the circular one. We can explain these results as follows: according to Petit \textit{et al.}, a higher intensity is reached in a filament with circularly polarized light because of a weaker multiphoton absorption. Consequently, beam self-focuses over a longer distance than for linear polarization \cite{Petit2000}. This leads to a smaller filament transverse size, as observed in figure \ref{fig_polarization}-(b). But since electron energy is higher in the case of circular polarization, heating is very efficient, which explains why a similar channel depth is reached in the two cases.

\section{Summary and conclusions}

We implemented a sensitive interferometric diagnostic for the characterization of the underdense channels generated by laser filamentation in the millijoule regime. We characterized the evolution of this channel over several orders of magnitude in time in the case of a \unit{5}{\milli\joule}, \unit{50}{\femto\second} laser pulse focused at $f/35$. These results are well reproduced by hydrodynamic simulations. We found that in these conditions, filamentation generates an underdense channel lasting for more than \unit{90}{\milli\second} and could result in the formation of a permanent density hole with laser repetition frequencies as low as \unit{10}{\hertz}. Moreover, numerical estimation of the initial air temperature following filamentation plasma recombination gives a value of \unit{1100}{\kelvin}, much higher than was previously reported in the case of low-energy filamentation.

We then investigated the influence of focusing conditions, of the wavelength and of the polarization of the laser pulse on the low-density channel. We were able to link the observed behaviors to the diffent sizes and densities of the plasma columns generated in each scenario. Energy transfer from the laser pulse to the medium thus appears optimized in case of strong focusing, with shorter wavelengths and using linear polarization, resulting in an important and spatially broad heating. We link underdense channel characteristics to the post-recombination temperature profile resulting from laser energy deposition in the medium. This opens perspectives for optimizing energy deposition in air by tailoring filamentation \cite{Panagiotopoulos2010,Suntsov2013}. According to the present analysis and results, the use of superfilaments \cite{Point2014b}, where the electron density is significantly higher than in standard multifilaments over several Rayleigh lengths, should result in an extended air column at very high temperature.

\acknowledgments{The authors thank Jérôme Carbonnel for technical assistance, and acknowledge financial support from the french DGA (grants n\textsuperscript{o} 2013.95.0901, n\textsuperscript{o} 2012.60.0011 and n\textsuperscript{o} 2012.60.0013).  Carles Mili\'{a}n and Arnaud Couairon thank Dr. Robin Huart for insightful discussions.}

\bibliographystyle{apsrev4-1}
\bibliography{biblio}

\end{document}